\documentclass{sf2a-conf}
\usepackage{graphicx}
\usepackage{natbib}
\begin{document}

\TitreGlobal{SF2A 2008}

\title{H$_{\textbf{2}}$ Energetics in Galaxy-wide Shocks \\ {\small Insights in Starburst Triggering and Galaxy Formation}}
\author{Guillard, P.}\address{Institut d'Astrophysique Spatiale, UMR 8617, CNRS, Universit\'{e} Paris-Sud 11, 91400 Orsay}
\author{Boulanger, F.$^1$}
\runningtitle{H$_{2}$ Energetics in Galaxy-wide Shocks: Insights in Starburst Triggering and Galaxy Formation}
\setcounter{page}{237}

\index{Guillard, P.}
\index{Boulanger, F.}

\maketitle
\begin{abstract} 

\textit{Spitzer} space telescope observations led to the surprising detection of a diverse set of extragalactic sources whose infrared spectra are dominated by line emission of molecular hydrogen. The absence or relative weakness of typical signs of star formation (like dust features, lines of ionized gas) suggest the presence of large quantities of $\rm H_2$ gas with no (or very little) associated star formation.
We use the Stephan's Quintet (SQ) galaxy collision to define a physical framework to describe the $\rm H_2$ formation and emission in galaxy-wide shocks. SQ observations show that exceptionally turbulent $\rm H_2$ gas is coexisting with a hot, X-ray emitting plasma. The extreme mid-IR $\rm H_2$ emission from the shock exceeds that of the X-rays.  These observations set a new light on the contribution of $\rm H_2$ to the cooling of the interstellar medium, on the relation between molecular gas and star formation, and on the energetics of galaxy formation.

These observations can be interpreted by considering that the shock is moving through an inhomogeneous medium. They suggest that most of the shock energy is transferred to bulk kinetic energy of the $\rm H_2$ gas. The turbulent energy of the post-shock gas drives a mass cycle across the different gas phases where $\rm H_2$ is forming out of the hot/warm gas. This interpretation puts the $\rm H_2$ emission into a  broader context including optical and X-ray observations. 
We propose that the turbulence in the clouds is powered by a slow energy and momentum transfer from the bulk motion of the gas and that the dissipation of this turbulent energy  in turn is powering the $\rm H_2$ emission.

\end{abstract}
%
\section{Introduction}
 
Recently, Spitzer IRS (Infra-Red Spectrometer) observations led to the unexpected detection of extremely bright mid-IR ($L > 10^{41}$erg~s$^{-1}$) $\rm H_2$ rotational line emission from warm gas towards the group-wide shock in Stephan's Quintet (hereafter SQ) (Appleton et al. 2006). 
This first result was quickly followed by the detection of bright $\rm H_2$ line emission from more distant galaxies (Egami et al. 2006, Ogle et al. 2007) and from the NGC 1275 and NGC 4696 cooling flows (Johnstone et al. 2007). These $\rm H_2$-bright galaxies may represent an important signature of galaxy evolution, but this unusual emission accompanied by little (or no) star formation has not yet been explained. 
Because of the absence or relative weakness of star forming signatures (dust features, ionized gas lines) in the mid-infrared Spitzer spectra, their exceptional H$_2$ luminosity may trace a {\it burst} of kinetic
energy dissipation  associated with galaxy interaction,
gas infall or AGN feedback. In \S~2 we briefly present the $\rm H_2$-luminous compilation of objects observed by \textit{ Spitzer}. In \S~3 we summarize the results obtained from our model (P. Guillard et al. 2008) of the SQ post-shock gas. Then \S~4 discuss the implications of these results on our understanding of the kinetic energy dissipation in these systems.

\section{An Emerging Population of H$_{2}$-Bright Sources}

\begin{figure}
   \centering
   \includegraphics[width=17cm]{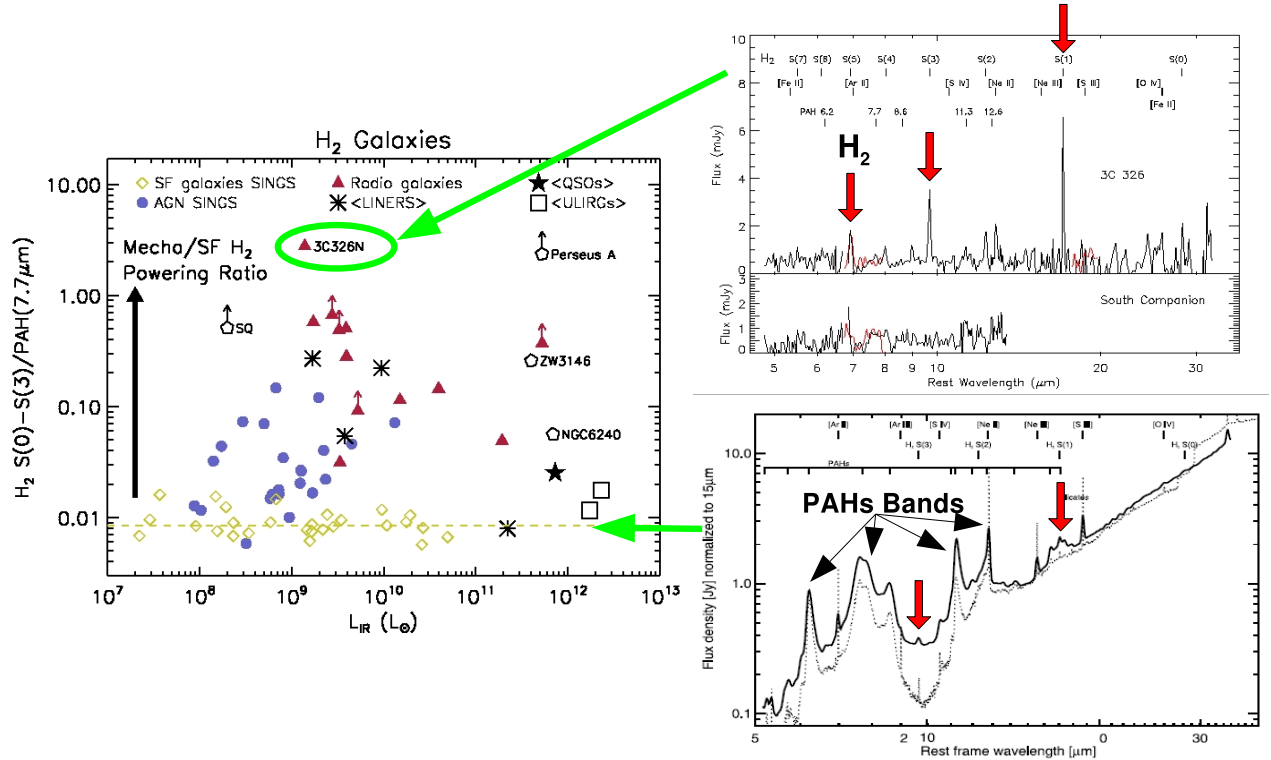}
      \caption{The H$_{2}$ rotational lines emission is compared with the PAH emission in the 7.7 micron band (left side). The large black symbols represent averages for samples of galaxies. The smaller symbols represent individual galaxies. At all IR luminosities, the observations reveal galaxies with excess H$_{2}$ emission on top of the Star Formation (SF) contribution defined by SF dominated galaxies (dashed line).   These H$_{2}$-Galaxies (MOHEGs) include active galactic nuclei galaxies (Seyferts, LINERs and radio galaxies), cooling flows (Perseus A, ZW3146) and colliding galaxies/mergers (SQ and NGC 6240). The excess H$_{2}$ emission reveals large (up to 10$^{10}$~M$_{\odot}$ ) quantities of warm ($T > 150$~K) molecular gas. On the right side we show for comparison an example of spectra of the 3C326 H$_{2}$-luminous radio-galaxy (from Ogle et al. 2007) with respect to normal star forming galaxies (adapted from Brandl et al. 2006)}
       \label{figure1}
   \end{figure}

Fig.~1 shows a sample of this new class of extremely luminous H$_{2}$
emission galaxies, up to 10$^{10}$~L$_{\odot}$ in pure rotational
molecular hydrogen emission lines and relatively weak total IR emission(Ogle et al. 2009, in prep.). 
The most contrasted examples show bright H$_{2}$ emission lines with no spectroscopic signature (dust or ionized gas lines) of star formation (Appleton et al. 2006, Ogle et al. 2007).
In many of these galaxies, molecular gas has been detected through the mid-IR
H$_{2}$ rotational lines prior to any CO observation. 
The same properties of the H$_{2}$ emission are observed in different types of objects that characterize
key-processes at work in galaxy formation and evolution: gas accretion, galaxy interactions, gas ejection due to star formation or to the action of the central black hole on its environment. In each case, H$_{2}$ lines appear to have an unexpected contribution to the gas cooling.

\section{The Stephan's Quintet: a ideal laboratory to study H$_{2}$ Energetics in Galaxy-Wide Shocks}

SQ is a nearby ($94$~Mpc) $\rm H_2$-luminous source where observations provide a clear astrophysical context to study the origin of the $\rm H_2$ emission. 
A wide ($5 \times 25$~kpc) shock is creating by a galaxy (NGC~7318b) colliding into a tidal tail at a relative velocity of $\sim 1\,000$~km~s$^{-1}$. Evidence for a galaxy-wide shock comes from observations of X-rays from the hot post-shock gas (Trinchieri et al. 1003, 2005), strong radio synchrotron emission from the radio emitting plasma (Sulentic et al. 2001) and shocked-gas excitation diagnostics from optical emission lines (Xu et al. 2003).
The surprise comes out from \textit{Spitzer} observations that show that this gas also contain molecular hydrogen with an $\rm H_2$ linewidth of 870 km~s$^{-1}$ (Appleton et al. 2006). The main energy reservoir is the bulk kinetic energy of the gas. A minor fraction of the collision energy is used to heat the hot plasma. The $\rm H_2$ surface brightness is larger than the X-ray emission from the same region, thus the $\rm H_2$ line emission dominates over X-ray cooling in the center of the shock. As such, it plays a major role in the energy dissipation and evolution of the post-shock gas. 

\begin{figure}
 \begin{minipage}[l]{.46\linewidth}
 \includegraphics[angle=90, width=9cm]{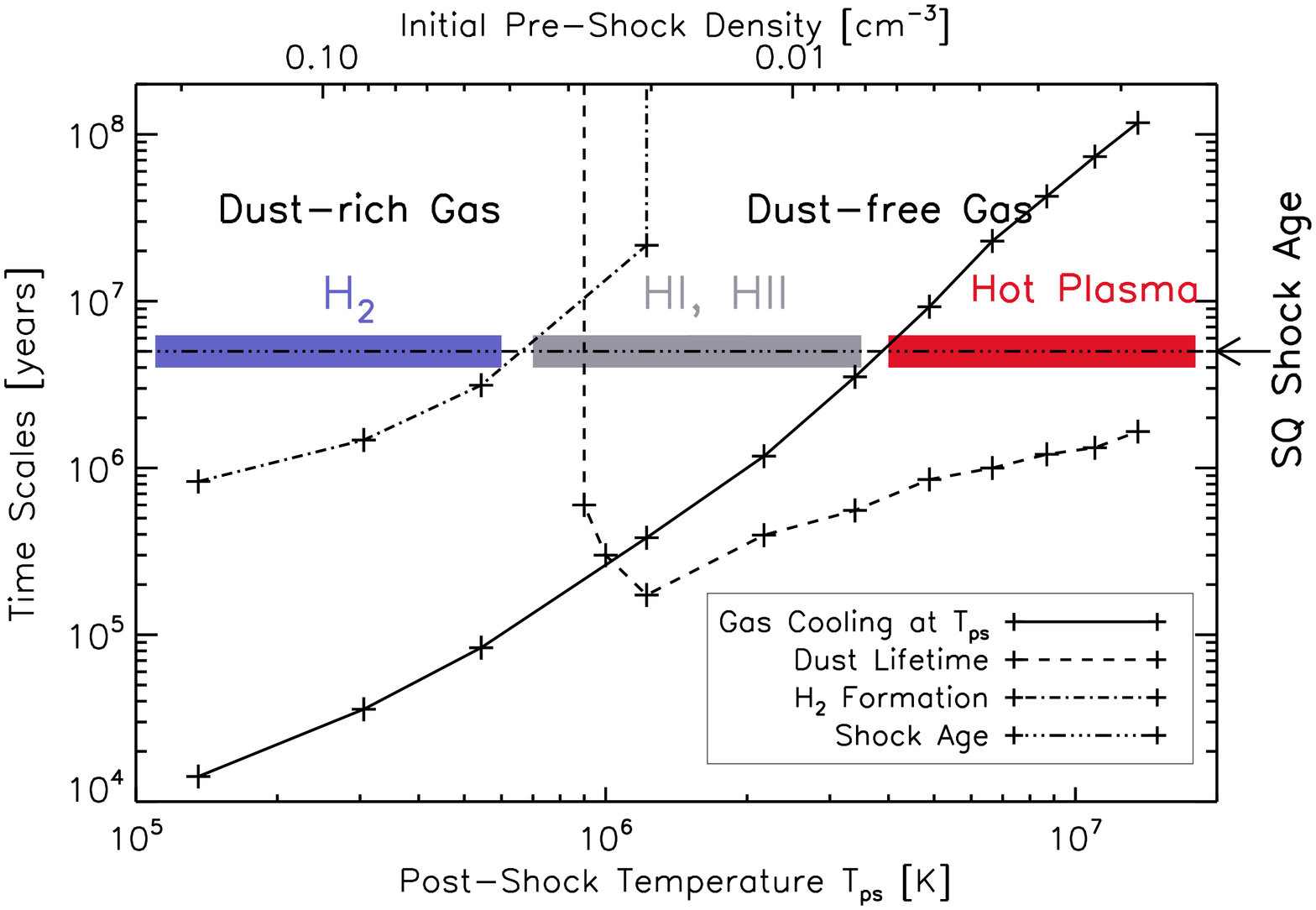}
 \caption{The multiphasic post-shock medium. The dust lifetime, $\rm H_2$ formation and gas cooling (full line) time scales in the SQ post-shock gas are plotted as a function of the post-shock temperature. Each shock velocity corresponds to a post-shock temperature or density. The ordering of the timescales separates the 3 main phases of the post-shock gas. The comparison of the $\rm H_2$ formation time scale with the SQ shock age ($\sim 5 \times 10^6\,$yr, indicated by the position of the bars) show where $\rm H_2$ molecules can form (blue bar). }
   \end{minipage} \hfill
   \begin{minipage}[r]{.46\linewidth}
       \includegraphics[width=7.6cm]{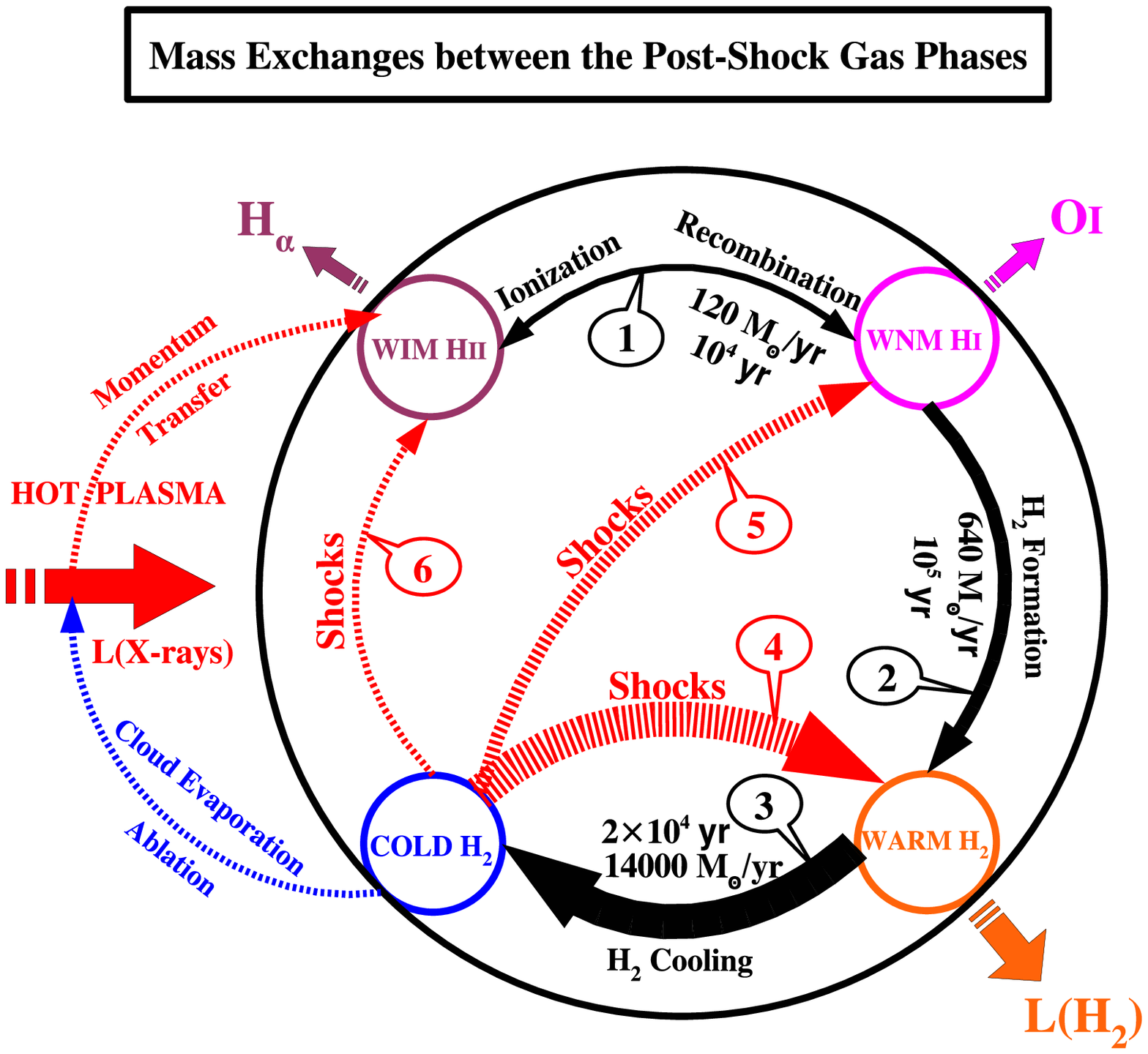}
   \caption{Schematic view at the gas evolutionary cycle proposed in our interpretation of Stephan's Quintet
optical and H$_2$ observations.
Arrows represent the mass flows between the H~II, warm H~I, warm and cold H$_2$ gas components. They are numbered for clarity.
The dynamical interaction between gas phases drives the cycle.
The mass flow values and associated timescales are derived from the $\rm H_{\alpha}$, O{\sc i}, and $\rm H_2$ luminosities and model calculations.
Heating of the cold H$_2$ gas (red arrows) is necessary to account for the increasing mass flow from the ionized gas to cold H$_2$  phases. }
   \end{minipage}
       \label{figure2}
   \end{figure}

We propose a scenario where a large-scale shock wave overtakes an inhomogeneous pre-shock medium (Guillard et al. 2008, submitted). The collision speed is the shock speed in the low density volume filling gas.
The post-shock pressure of the hot gas drives slower shocks into denser gas. The post-shock gas is thus heated to a range of temperatures that depend on the pre-shock gas density. Our model quantifies the gas cooling, dust destruction, H$_{2}$ formation and excitation in the post-shock medium (Fig.~2).

Schematically, the low density volume filling pre-shock gas, shocked at high velocities ($\sim~600\,$km~s$^{-1}$) becomes a dust-free X-ray emitting plasma. The cooling timescale of the hot gas is more than one order of magnitude greater than the shock age ($5 \times 10^{6} \rm \ yr \ $). Therefore the post-shock plasma does not have the time to cool down and form molecular gas since the shock was initiated. Denser gas is heated at lower temperatures and dust survives. In the context of the SQ shock, we show that these clouds have time to cool down and form H$_{2}$ before being disrupted. 
We show that the~cooling of the H{\sc ii} or H{\sc i} gas cannot explain the observed H$_{2}$ emission, but that low-velocity ($5-20$~km~s$^{-1}$) shocks driven into molecular fragments can account for it. We propose that these shocks are generated by cloud fragments collisions. The supersonic turbulence in the molecular fragments is powered by a slow energy and momentum transfer from the bulk motion of the gas.
The turbulent energy of the post-shock gas drives a mass cycle across the different gas phases where H$_{2}$ is forming out of the hot gas.

\section{The Cycling of Gas across ISM Phases}

In this section, we place the Spitzer $\rm H_2$ detection in the broader context set by optical observations. We propose a picture of the post-shock gas evolution that sketches the dynamical interaction between gas phases. This interpretation introduces a physical framework that may apply to H$_2$ luminous galaxies in general.

A schematic cartoon of the evolutionary picture of the post-shock gas which arises from our data interpretation is presented in Fig.~3. It sketches a global view that we detail here. Black and red arrows represent the mass flows between the H~II, warm H~I, warm and cold H$_2$ gas components of the post-shock gas. The large arrow to the left symbolizes the relative motion between the warm and cold gas and the surrounding plasma.
Each of the black arrows is labeled with its main associated process: gas recombination and ionization (arrow number 1), H$_2$ formation (2) and H$_2$ cooling (3). The values of the mass flows and the associated timescales are derived from observations and our model calculations (see Guillard et al. 2008 for details).

A continuous cycle through gas components is excluded by the increasing mass flow
needed to account for the $\rm H_{\alpha}$, O{\sc i}, and $\rm H_2$ luminosities.
Heating of the cold H$_2$ gas towards warmer gas states (red arrows) needs to occur.
It is the dissipation of the gas mechanical energy that powers these red arrows.
The post-shock molecular cloud fragments are likely to experience a distribution of shock velocities, depending on their size and density. Arrow number 4 represents the low velocity magnetic shocks excitation of H$_2$ gas that can account for the H$_2$ emission (described in Guillard et al.).
More energetic shocks may dissociate the molecular gas (arrows number 5).
They are necessary to account for the low H$ _{\alpha}$ to O~{\sc i} luminosity ratio.
Even more energetic shocks may ionize the molecular gas (arrow number 6).
This would bring cold $\rm H_2$ directly into the H~{\sc ii} reservoir.

Turbulent mixing of hot, warm and cold
gas are an alternative mass input of H~{\sc ii} gas.
Turbulent mixing layers result from cloud gas shredding into fragments
that become too small to survive evaporation due to heat conduction (see e.g. Begelman \& Fabian, 1990).
Turbulent mixing is represented by the thin blue arrow towards the surrouding plasma,
to the left of our cartoon. Turbulent mixing produces intermediate temperature gas that is thermally instable. This gas cools back to produce H~II gas that enters the cycle (thin red arrow). It is relevant to our scenario to note that cold gas in mixing layers probably preserves its dust content. It is only heated to a few $10^5\,$K, well below temperatures for which thermal sputtering becomes effective. Further, metals from the dust-free hot plasma that is brought to cool are expected to accrete on dust when gas cools and condenses.

\section{Concluding remarks}

Our understanding of the dynamical interaction between gas phase is inspired by
numerical simulations (e.g. Audit \& Hennebelle 2005)
which have changed our perspective on the interstellar medium phases from a static to a dynamical picture. The interaction with the hot plasma supplies mechanical energy and momentum to the warm gas as discussed in the context of cold gas observations in clusters by Pope et al. 2008. In our interpretation it is this energy input which drives turbulence and the gas cycle.

Many galaxy collisions and mergers are observed to trigger IR-luminous bursts of star formation. However, the absence of spectroscopic signatures of photoionization (dust or ionized gas lines) in the center of the SQ shock (Xu et al. 2003) show no or little star formation  in this region. Our interpretation is that the bulk kinetic energy of the gas colliding flows  is not completely dissipated. Within this dynamical understanding of the post-shock gas, cold molecular is not a mass sink. The fact that there is no star formation at the centre of the shock shows that it is not long-lived enough to allow formation of gravitationally instable molecular fragments. Within a more general framework, interacting H$_2$ luminous galaxies may represent an intermediate phase in the evolution of mergers, prior to the starburst.


\end{document}